\documentclass{emulateapj}
\def\mathbi#1{\textbf{\em #1}}
\let\oldhat\hat
\renewcommand{\vec}[1]{\mathbi{#1}}
\renewcommand{\hat}[1]{\oldhat{\mathbf{#1}}}

\usepackage{graphicx}
\usepackage{subfigure}
\usepackage{epsf}
\usepackage{longtable}

\begin{document}

\title{The Origin and Distribution of Cold Gas in the Halo of a Milky Way-Mass Galaxy}

\author{Ximena Fern\'andez, M. Ryan Joung \& Mary E. Putman}
\affil{Department of Astronomy, Columbia University, 550 West 120th Street, New York, NY 10027}

\begin{abstract}
We analyze an adaptive mesh refinement hydrodynamic cosmological simulation of a Milky Way-sized galaxy to study the cold gas in the halo.  HI observations of the Milky Way and other nearby spirals have revealed the presence of such gas in the form of clouds and other extended structures, which indicates on-going accretion.  We use a high-resolution simulation (136-272 pc throughout) to study the distribution of cold gas in the halo, compare it with observations, and examine its origin.  The amount ($\sim 10^8~$ M$_\odot$ in HI), covering fraction, and spatial distribution of the cold halo gas around the simulated galaxy at $z=0$ are consistent with existing observations.  At $z=0$ the HI mass accretion rate onto the disk is 0.2 M$_\odot$ yr$^{-1}$.  We track the histories of the 20 satellites that are detected in HI in the redshift interval $0.5>z>0$ and find that most of them are losing gas, with a median mass loss rate per satellite of $3.1 \times 10^{-3}$ M$_\odot$ yr$^{-1}$.  This stripped gas is a significant component of the HI gas seen in the simulation. In addition, we see filamentary material coming into the halo from the IGM at all redshifts.  Most of this gas does not make it directly to the disk, but part of the gas in these structures is able to cool and form clouds.  The metallicity of the gas allows us to distinguish between filamentary flows and satellite gas. We find that the former accounts for at least $25-75\%$ of the cold gas in the halo seen at any redshift analyzed here.  
Placing constraints on cloud formation mechanisms allows us to better understand how galaxies accrete gas and fuel star formation at $z=0$.
\end{abstract}

\section{Introduction}
Determining how galaxies accrete their baryons is of crucial importance to understand how galaxies grow and evolve. The Milky Way has had a nearly constant star formation rate over the past 5--7 billions of years \citep{Binney00}, implying there must be a continuous replenishment of gas.  In addition to this, the metallicity distribution of stars in the Galaxy would be better explained if there is a fresh supply of gas \citep[e.g.,][]{Larson72,Chiapini01}.  This on-going accretion is in contradiction with the classical models of galaxy formation where all of the gas is shock heated to the virial temperature and monolithically collapses within a cooling radius \citep*[e.g.,][]{White78, WhiteFrenk91}.  

Both simulations and observations provide evidence for gradual accretion from the baryonic halo onto the disk of galaxies.  Hydrodynamic simulations have shown that galaxies can accrete gas directly from the intergalactic medium (IGM).  In low-mass galaxies ($M_{\rm gal} \lesssim 10^{10.3}$ M$_{\odot}$) at high redshifts,  most of the gas never shock-heats to the virial temperature and is delivered to galaxies through filaments in a cold mode \citep[e.g.,][]{KeresKatz05}.  For high-mass galaxies at low redshift, cold mode accretion is no longer dominant since filaments get disrupted and are not able to reach the disk.  \citet[][hereafter KH09]{Keres09} show in their SPH simulation of a Milky Way-mass galaxy that the gas from these disrupted filaments is still able to reach the galaxy.  The filaments trigger instabilities in the halo, which lead to the formation of clouds.  Cold gas in the halo is also seen in other simulations, where cooling instabilities in the hot halo condense out into warm clouds \citep{Kaufman06, SL06}.  This idea has been challenged by recent analytic calculations showing that clouds will not develop from thermal instabilities  \citep{Binney09}. Instead, nonlinear perturbations, such as those seeded by filamentary flows, could lead to the formation of clouds \citep{Joung11}.   These clouds might reach the disk of the galaxy and provide continuous fuel for star formation.  

There is robust observational evidence for the presence of cold gas in the baryonic halo of spiral galaxies. In our own Galaxy, there is a population of high-velocity clouds (HVCs) that reside outside of the disk, whose kinematics indicate a net infall.  These were first detected in neutral hydrogen (HI) by \citet{Muller63}, and since then Galactic HI surveys have provided detailed maps of the distribution of the HVC population \citep[e.g.,][]{Putman02, Peek11}, and distance constraints have placed the clouds within the Galactic halo \citep{Putman03,Thom06,Wakker08}.  HI observations of nearby spirals have revealed analogs to these HVCs \citep[see][for a review]{Sancisi08}.  The extraplanar gas in M31 has been extensively studied since it is a nearby large spiral \citep[e.g.,][]{Thilker04, Westmeier08}, and new deep HI surveys are being carried out to detect this low-density gas in a larger sample of galaxies \citep[e.g.,][]{Heald11}.  In addition to HI observations, absorption-line studies have revealed the presence of lower density gas in the halos of galaxies at low and intermediate redshift \citep[e.g.,][]{Wakker09,Chen10}. 

There are different scenarios that could explain the origin of the observed cold gas in galaxy halos.   These scenarios can generally be divided into two groups: (1) satellite material, and (2) cooling over-densities within the hot halo. In the first scenario, some of the gas is likely to be a consequence of the interaction of satellites with the hot halo.  Studies of this show that satellites lose their gas due to ram pressure stripping, and some of this stripped gas may resemble cold halo gas features \citep{Mayer06,Grcevich09,Nichols11}.  In the second scenario, the cooling over-densities could originate from filamentary flows or galactic fountains.  Filamentary material has been seen in simulations by KH09, where they find that the clouds are created via cooling and Rayleigh-Taylor instabilities caused by cold flows generating density inversions in the hot medium.   As an alternative, some of the cooling gas in the halo might have originated from the disk itself via supernovae outflows \citep{Shapiro76,Bregman80}. This, however, could only account for some of the cold gas since many HVCs are found at large distances with low 
(0.1--0.3 Z$_{\odot}$) 
metallicities \citep[e.g.,][]{Wakker01}.     

In this paper, we analyze an adaptive mesh refinement hydrodynamic cosmological simulation of a Milky Way-sized galaxy to understand how spiral galaxies get their gas.  We mostly study HI because of the large amount of observational data available on the spatial distribution and kinematics of this gas component.  We focus on stripped gas and filamentary flows as possible origins of the cold halo gas. Throughout our discussion we use the expressions  ``filamentary flows" and  ``cold flows" interchangeably  to describe filamentary gas in the halo that has never been shock-heated to the virial temperature.

The structure of the paper is as follows. The simulation we analyze is described in detail in \S 2.  In \S 3,  we present our findings, which includes the properties of the cold gas around the simulated galaxy at $z=0$, a calculation of the HI mass accretion rate, and an investigation of the origin of the gas.  We then compare the simulation with observations of M31 and other galaxies in \S 4, and examine the HI covering fraction.  We discuss the implications of our results in \S 5 and conclude in \S 6.   

\section{Simulation}
We analyze a cosmological simulation (Joung et al., in prep) 
performed with \textit{Enzo}, an Eulerian hydrodynamics code with adaptive mesh refinement (AMR) capability  \citep{Bryan99,Norman99,Oshea04}.  It solves the Euler equations using the piecewise-parabolic method \citep*[PPM;][]{Colella84} or the hydro solver used in Zeus \citep*{Stone92} to handle compressible flows with shocks; the latter was used primarily for numerical stability.

First, a low-resolution simulation was run with a periodic box of $L =$ 25 $h^{-1}$ Mpc comoving on a side with cosmological parameters consistent with WMAP5 \citep[][; $\Omega_m$, $\Omega_{\Lambda}$, $\Omega_b$, $h$, $\sigma_8$,$n_s$) $=$ (0.279, 0.721, 0.046, 0.70, 0.82, 0.96)]{Komatsu09}.  Local Group-like volumes were identified based on the halo mass (mass range 1--2 $\times$ 10$^{12}$ M$_{\odot}$), the mean density (0.60--1.0 times the mean density of the universe) and the relatively low velocity dispersion of the halos ($<$ 200 km s$^{-1}$) identified within 5 $h^{-1}$ Mpc of a given galaxy.  Four halos met such criteria.  A resimulation for one of the four halos was done using the multimass initialization technique \citep{Katz94,Navarro94} with four nested levels (five including the root grid), achieving $m_{DM} = 1.7 \times 10^5$ M$_{\odot}$, within a ($\sim$5$h^{-1}$ Mpc)$^3$ subvolume.  The selected galaxy has a halo mass of $1.4\times 10^{12}$ M$_{\odot}$ at $z=0$ and so contains over 8.2 million dark matter particles within the virial radius.  With a maximum of 10 levels of refinement, the maximum spatial resolution stays at 136--272 pc comoving or better at all times.

The simulation includes metallicity-dependent cooling \citep{Cen95} extended down to 10 K \citep*{Dalgarno72},  metagalactic UV background \citep{Haardt96}, shielding of UV radiation by neutral hydrogen \citep{Cen05}, and a diffuse form of photoelectric heating \citep{Abbott82,Joung06}. The code simultaneously solves a complex chemical network involving multiple species (e.g., HI, HII, H$_2$, HeI, HeII, HeIII, e$^{-}$; \citealt{Zhang}) 
and metal densities explicitly.

Star formation and stellar feedback, with a minimum initial star particle mass of $m_* = 1.0 \times 10^5$ M$_{\odot}$, are also included.  Star particles are created in cells that have $\rho > \rho_{SF}$, where $\rho_{SF}=7 \times 10^{-26}$ cm$^{-3}$ \citep{Navarro93,Naab07} and that do not satisfy the Truelove criterion for stability \citep{Truelove97}. The star formation efficiency, defined as the fraction of gaseous mass converted to stars per dynamical time,  is 0.03.  Each star particle is tagged with its mass, creation time, metallicity, and a unique identification number.  The metal enrichment inside galaxies and in the IGM is followed self-consistently.  Supernovae feedback 
is modeled following \citet{Cen05} with $e_{SN} = 10^{-5}$.  Feedback energy and ejected metals are distributed into 27 local cells centered at the star particle in question, weighted by specific volume of the cell.  The temporal release of metal-enriched gas and thermal energy at time $t$ has the following form: $f(t, t_i, t_{\rm dyn}) = (1/t_{\rm dyn})[(t-t_i)/t_{\rm dyn}] \exp[-(t-t_i)/t_{\rm dyn}]$, where $t_i$ is the formation time of a given star particle.  The metal enrichment inside galaxies and in the IGM is followed self-consistently in a spatially resolved fashion.  We identify virialized objects in our high-resolution simulation using the HOP algorithm \citep*{Eisenstein98}.

\begin{figure*}
\begin{center}
\includegraphics[trim= 0 40 0 0,clip,scale=.7]{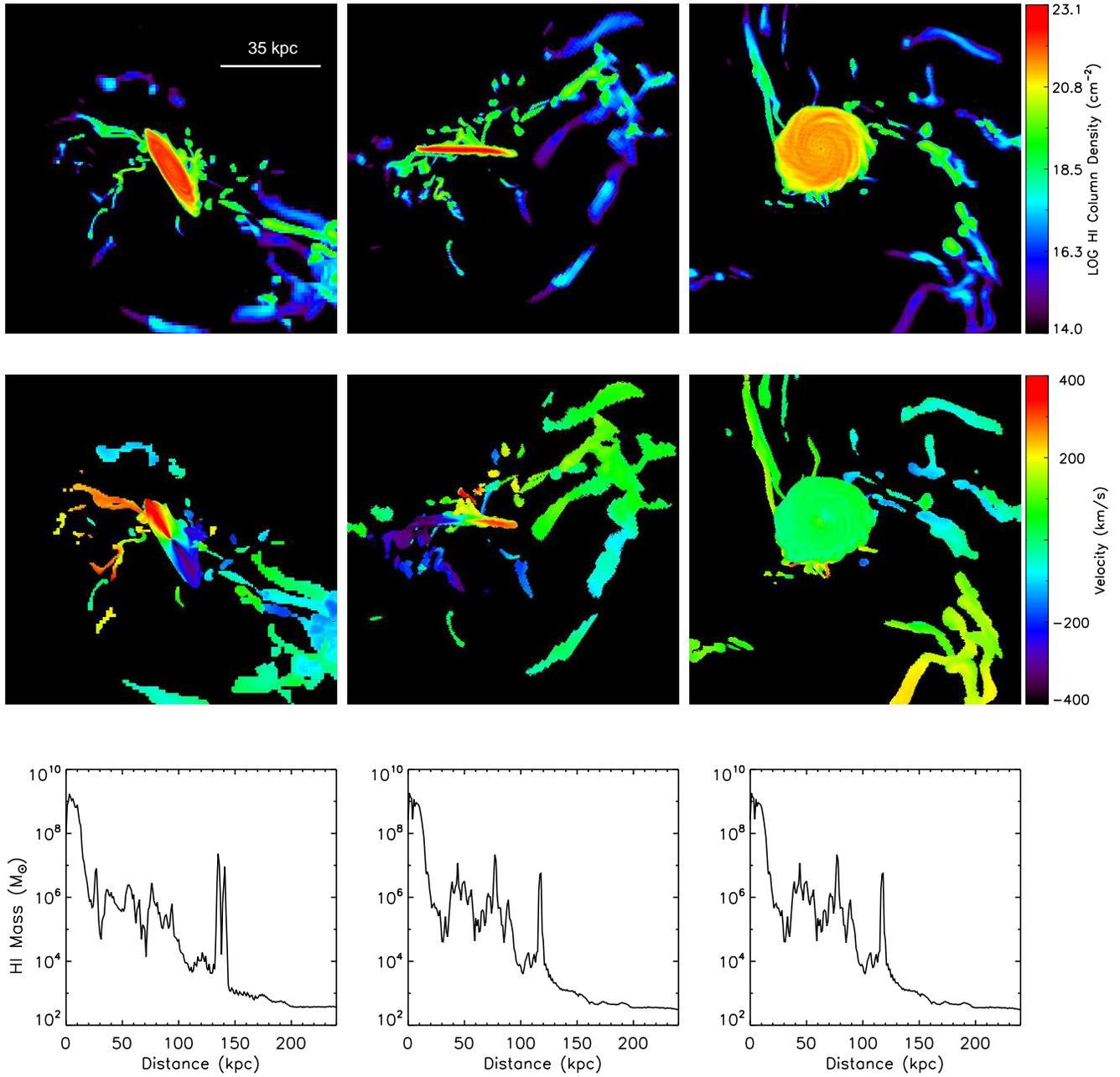}\\
\caption{Panels showing 3 different projections: first column shows the unrotated simulated galaxy, the second one shows an edge-on view, and the third shows the galaxy face-on.  The different rows show the HI column density map (first), velocity map (second), and the HI mass as a function of distance. 
Both the HI column density and velocity maps have a cutoff of $10^{14}$ cm$^{-2}$ in column density; lines of sight with lower column densities were masked in black. The HI mass distribution in the third row is calculated using annuli of 1 kpc thickness.}
\label{fig:panelsObs}
\end{center}
\end{figure*}

\section{Results and Analysis}

\subsection{Properties of the Simulation at z=0}
We first study the HI distribution and kinematics at $z=0$ viewed from different angles.  We construct Position-Position-Velocity (PPV) cubes from the HI density and velocity arrays for three projections: the unrotated frame\footnote{The unrotated frame is along the $x$-axis of the computation box, with $i=34^\circ$ and PA=$  109^\circ$}, edge-on, and face-on.  We do this by first creating a velocity range with a 20 km s$^{-1}$ resolution that includes all of the velocities seen in the simulation, 
and then assigning the simulation cells to a specific bin depending on their line-of-sight velocity (which depends on the projection). In each velocity bin, we compute a column density by adding along the line-of-sight, which results in a HI density map for positions $(x,y)$ 
at different velocities.   We then take the first and second moments along the velocity axis to make HI column density and velocity maps. We show these maps in Figure~\ref{fig:panelsObs} with a field of view of $108 \times 108$ kpc$^2$.  In addition to the maps, we calculate the amount of HI as a function of projected distance from the center of the galaxy.  The different columns in Figure~\ref{fig:panelsObs} correspond to the three viewing angles.  The plots in the first two rows are made using the highest-resolution extraction box (ten levels of refinement; 1 cell size=0.27 kpc), while the plots in the bottom row were produced from eight levels of refinement (1 cell size=1.09 kpc) which provides a larger field of view (480 kpc on a side).  

\begin{figure*}
\begin{center}
\hspace{0.2in}
\includegraphics[trim= 20 45 10 40,clip,scale=0.8]{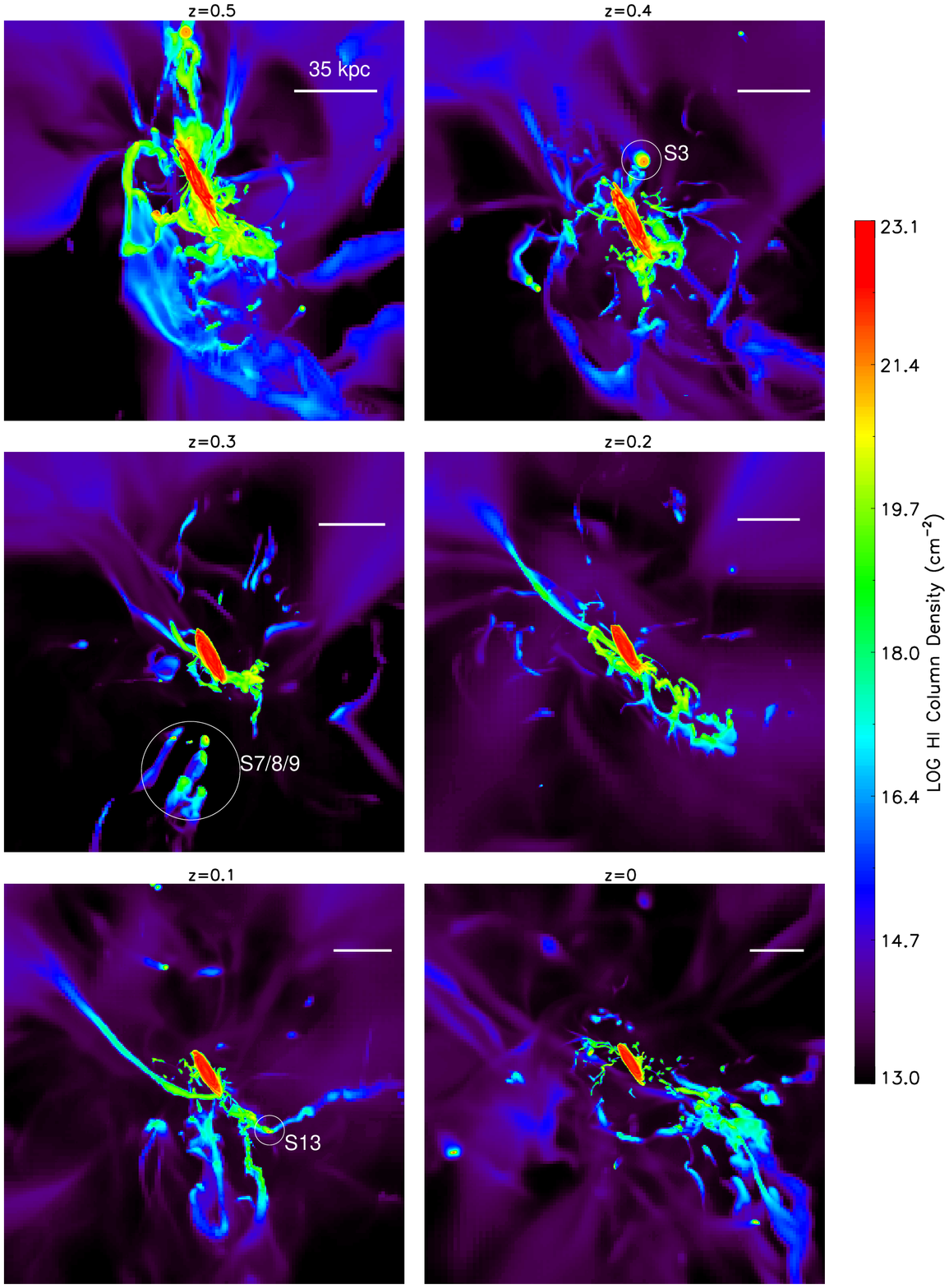}
\caption{Snapshots showing the HI distribution at six different redshifts above a column density of $10^{14}$ cm$^{-2}$.  The horizontal line in the upper right-hand corner of each panel corresponds to 35 kpc physical, while the field of view is fixed at 270 kpc comoving on a side.  We circle  examples of satellites interacting with the main galaxy (S3), with one another (S7/8/9), and with the hot halo gas (S13).  These panels also show that cold flows provide an important fueling mechanism at all times.}
\label{fig:HIMaps}
\end{center}
\end{figure*}

The first row corresponds to HI column density maps, which include column densities above $10^{14}$ cm$^{-2}$.  The disk of the simulated galaxy spans about 15-20 kpc in radius, and the halo population exhibits a wide array of morphological features that tend to be arranged in complexes.  We see compact clouds, elongated features resembling filaments, and interconnected structures.  Most of these structures are located in the vicinity of the disk, and tend to have higher column densities than the gas seen at larger distances.  The features seen at the edges of the figures are part of bigger complexes (see Figure~\ref{fig:HIMaps} for a bigger field of view at $z=0$).  The edge-on view depicts a small warp to the right of the disk, which interestingly points in the direction of stripped gas material that is located below the disk.  This is suggestive of the warp being caused by the satellite as it approached the disk.  The face-on projection shows that many of these clouds were once part of coherent structures that fragmented as they approached the disk.  
\begin{figure*}
\begin{center}
$\begin{array}{cc}
\includegraphics[trim= 30 10 50 0,clip,scale=0.6]{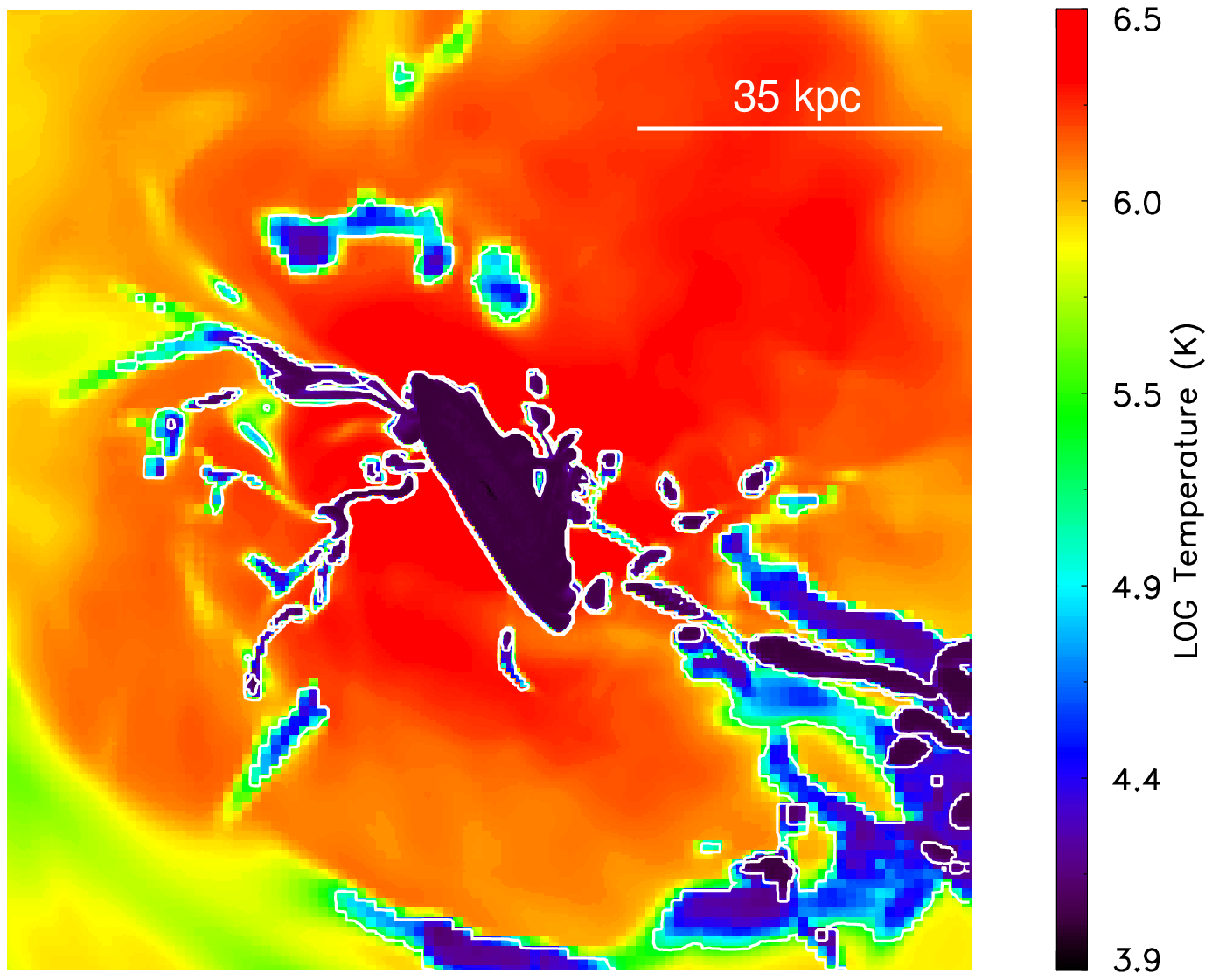}&
\includegraphics[trim= 50 20 50 0,clip,scale=0.63]{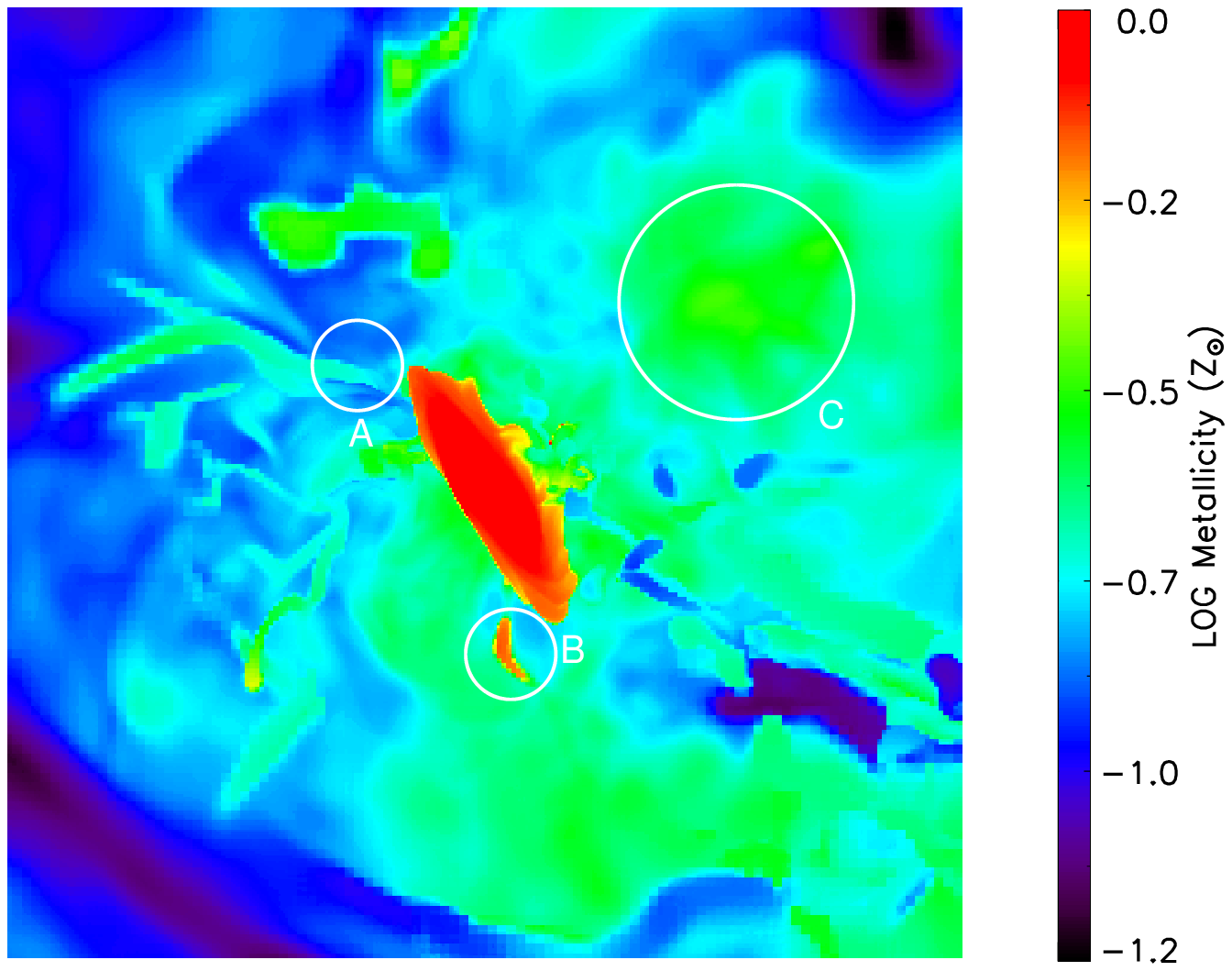}\\
\end{array}
$
\caption{Temperature and metallicity map both weighted by HI density along each line-of-sight.  The projection is the same as that used for the first column in Figure 1. We overlay HI contours of 
0.0001 and 1 
$\times 10^{18}$ cm$^{-2}$ on the temperature map.  In the metallicity plot, we enclose three features that show the different metallicities.  Circle A is a filamentary flow, Circle B shows the stripped gas from a satellite, and Circle C is 
caused by 
supernova feedback.  Note that the satellite and supernova feedback gas are highly enriched, while the filamentary flow is associated with lower metallicities.}
\label{fig:metTemp}
\end{center}
\end{figure*}

The velocity maps presented in the second row use the line-of-sight velocity for direct comparison with observations. We only include velocity cells with a HI column density above $10^{14}$ cm$^{-2}$ when making the map.  We subtract the systemic velocity of the galaxy, and plot velocities within the range between $-400$ and 400 km s$^{-1}$.  The kinematics of the clouds overall follow the rotation of the disk.  The big complex seen in the bottom right-hand corner of the unrotated view is at the systemic velocity, indicating that the gas is not moving significantly with respect to the galaxy.  The gas in the immediate vicinity of the disk has velocities indicating infall.  We also see the warp in the kinematics of the edge-on view, but most of the HI gas in its proximity is near the systemic velocity, not showing a direct connection to the warp. 
The face-on projection does not show any extreme behavior in the clouds velocity, since most of the visible gas in that projection is near the systemic velocity. Overall, the three viewing angles show that most of the cold halo gas structures have velocities near the systemic velocity, or have velocities that indicate infall. This is consistent with the 
appearance of gas motions in a movie made from multiple timesteps in the simulation, 
where we see gas accreting onto the disk.  

The plots in the third row show how the HI is distributed as a function of projected distance from the center of the galaxy. We created these plots by constructing spherical annuli of 1 kpc in thickness with their origin at the center of the galaxy and calculating the amount of HI in each. We calculate the HI mass out to 240 kpc by using the level 8 refinement extraction box without imposing a column density limit.  The distribution is overall declining with distance, with the first 20 kpc corresponding to the gas found in the disk.  Note that the HI gas density within the solar neighborhood is consistent with the observed value of 13.2 M$_\odot$ pc$^{-2}$ \citep{Flynn06}. There is at least $10^5$ M$_\odot$ of HI gas in each annulus
out to 90 kpc, where there is a peak at around 40-50 kpc, which is due to the complex seen at the edges of the column density maps in the first row. After this, the HI mass distribution continues to decrease to values that are not observable, with isolated spikes that correspond to satellites.

Figure~\ref{fig:metTemp} shows the temperature and metallicity maps that serve to further illustrate the properties of the HI at $z=0$.  Both maps are of the unrotated galaxy, and are HI-gas density weighted along each line-of-sight.  As seen in the temperature  map, the galaxy is embedded in a diffuse hot halo of $1-3 \times 10^6$ K, whose mean temperature declines slowly with increasing distance from the galactic center, and is surrounded by colder fragmented flows with a range of temperatures $8 \times 10^{3} - 3 \times 10^{5}$ K.  The higher than expected HI gas temperatures are due to the weighting 
by density
along the line of sight.   We overlay HI contours of $10^{14}$  and  $ 10^{18}$ cm$^{-2}$ to show which structures are observable either in HI emission or absorption studies. 

\begin{figure}
\begin{center}
\includegraphics[trim= 25 0 10 0,clip,scale=0.52]{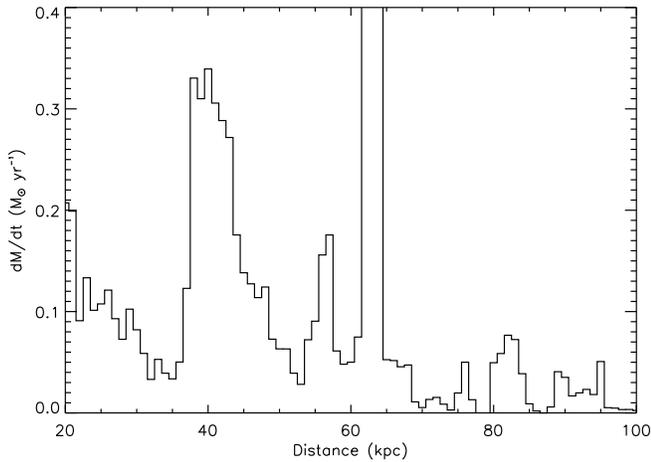}
\caption{HI mass accretion rate as a function of radius calculated using the level 8 extraction box (resolution of 1.09 kpc per cellsize).  We plot $\dot{M}_{\odot}$ for distances between 20 and 100 kpc, since the first 20 kpc correspond to material in the disk, and we do not see any significant HI at larger distances.  We calculate a mass accretion rate onto the disk of $\sim 0.2$ M$_{\odot}$  yr$^{-1}$, which is the value calculated at 20 kpc.  The HI seen at 40 kpc corresponds to a complex of cold gas moving at the systemic velocity, and the spike seen at 
63 
kpc is from a satellite that still has gas at $z=0$.
}
\label{fig:dmdt}
\end{center}
\end{figure}

The metallicity map shows that all of the gas in the halo has been enriched to some extent.  Most of the cold halo gas has an approximate metallicity of 0.15~Z$_\odot$, in contrast with the gas in the disk that has a metallicity of $\sim$1 Z$_\odot$.   The filaments show a slight level of enrichment ($Z=0.2$ Z$_\odot$), as seen in the feature enclosed by Circle A.  Another feature in the vicinity of the disk (Circle B) has a metallicity of 0.8 Z$_\odot$, corresponding to stripped gas from a satellite.  We can also see an enriched component of around 0.3 Z$_\odot$ perpendicular to the major axis of the disk, which corresponds to traces of the supernova feedback wind (Circle C).   As we will discuss later, the metallicity of the gas turns out to be an important indicator of the origin of the neutral gas.

\subsection{HI Mass Accretion Rate}
We calculate the HI mass accretion rate at $z=0$ as a function of radius from the level 8 extraction box following \citet{Peek08}.  We use the following equation: 

\begin{equation}
\dot{M}(R )=\sum_{i=1}^{n(R )} \frac{M_i\vec{V}_i \cdot (-\hat{\vec{r}}_i)} {dR}
\end{equation}

\noindent
where $M_i$ is the HI mass  in the $i^{th}$ cell, $\vec{V}_i$ is the velocity vector of that cell, $\hat{\vec{r}}_i$ is the radial unit vector, and $dR$ is the thickness of the spherical shell.  Figure~\ref{fig:dmdt} shows the resulting mass accretion rate as a function of radius.  We calculate the mass accretion rate out to 240 kpc, but only plot the first 100 kpc since there is no halo 
HI gas at larger distances. We exclude the first 20 kpc since that is the radius of the disk. As seen in the plot, the mass accretion rate does not linearly decrease with increasing radius.  The spike at 
63 kpc corresponds to one of the satellites that still has gas at $z=0$. Excluding this, there are various peaks at different distances. The two 
most massive 
ones are at 20 kpc and 40 kpc, the first one being due to the HI in the immediate vicinity of the disk, while the other one represents the big complex that is moving at the systemic velocity of the galaxy.  From this analysis, we estimate 
an HI 
mass accretion of $\sim 0.2$ M$_{\odot}$ yr$^{-1}$ onto the disk. 

\subsection{Origin of the Gas}
We now examine the origin of the neutral gas in the halo by analyzing a sequence of 117 projection maps taken from a range of redshifts each separated uniformly by 43 Myr, starting at $z=0.5$ and ending at $z=0$.   
This time sequence is of the unrotated galaxy, which gives a close to edge-on view with a larger field of view (540 kpc comoving on a side). Figure~\ref{fig:HIMaps} shows six representative snapshots at  $z=0.5, 0.4, 0.3, 0.2, 0.1$,and 0.0,  showing the HI distribution.  These maps show the presence of filamentary flows at all times, and also provide some examples of satellites getting stripped as they approach the main galaxy.  

\begin{figure}
\begin{center}
\includegraphics[trim= 25 0 10 0,clip,scale=0.52]{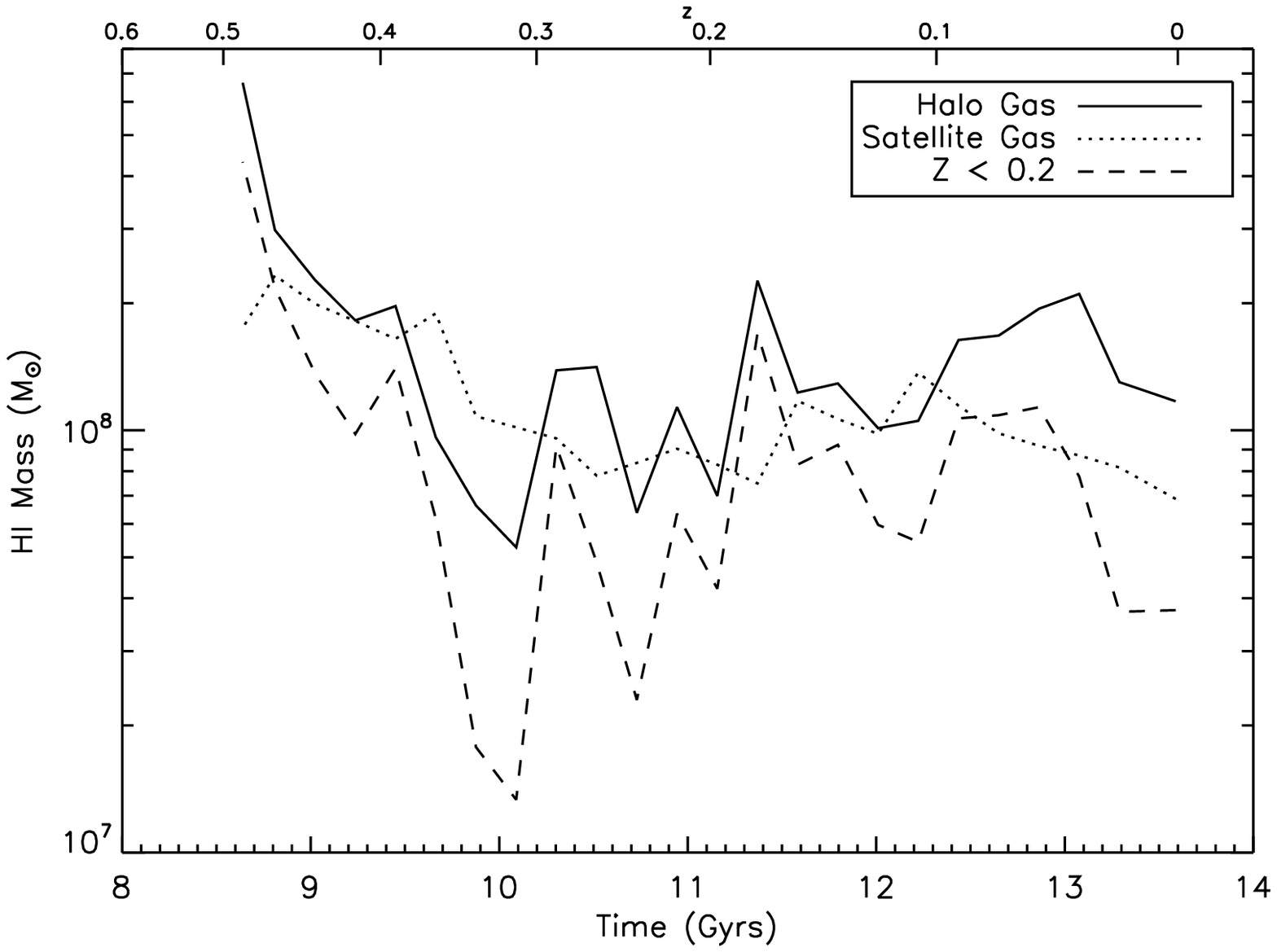}
\caption{Halo HI gas evolution between $z=0.5$ to $z=0$.  The  solid curve represents the HI gas at  a specific redshift (excluding gas in the disks and satellite cores), 
the dotted
curve represents gas inside the satellite cores, and the dashed one shows the low-metallicity gas which is largely representative of filamentary flows. We compute these three quantities at every five simulation 
outputs, corresponding to an interval of 215 Myr, without setting a column density limit. 
Note that the HI in the halo is roughly constant  ($\sim10^{8}$ M$_\odot$) since $z=0.3$.  The shape of the solid curve is closely followed by the low-metallicity gas, indicating the variations are due to fluctuations in filamentary accretion.}
\label{fig:evolution}
\end{center}
\end{figure}

In addition, Figure~\ref{fig:evolution} shows the evolution of the HI halo gas mass in the redshift interval $z=0.5-0.0$.  We calculate the amount at every five simulation outputs, which corresponds to an interval of 215 Myr.  In calculating this number, we do not set a minimum column density and exclude the amount of gas inside the main disk and in the satellite cores, which then leaves a combination of gas from filamentary flows and stripped galaxies. We remove the disk and satellites by excluding the high-column density gas ($ > 10^{20}$ cm$^{-2}$) within a specified distance from the center of the dark matter halo. The solid curve shows that the amount of HI is larger at earlier times, 
when 
the disk is actively accreting greater amounts of gas ($z>0.3$). At later times, the plot shows a roughly constant amount of HI gas in the halo ($\sim 10^{8}$ M$_\odot$.) There are some peaks at different times, which could be due to more satellites being stripped or to streams being more active at certain times.    We discuss how we studied these two possible origins for the cold halo gas and our findings below.  

\subsubsection{Satellites}

We identify 20 gas-rich satellites by selecting dark matter subhalos that have both stars and HI gas (above $10^{16}$ cm$^{-2}$). 
Figure~\ref{fig:HIMaps} provides examples of  what happens to satellites as they approach the main galaxy.  In the second snapshot (at $z=0.4$) there is a spherical satellite immediately above the disk, which we label S3.  This satellite eventually merges with the disk, and has no noticeable tidal tails in its HI morphology in this projection.  At $z=0.3$, we see another satellite approaching the main galaxy from the bottom left
corner.  This is actually a system of three galaxies (S7, S8, S9), and its HI morphology is visibly disturbed, as evident from the elongated features.  This is possibly a product of both tidal interactions within the system and 
an 
interaction with the hot medium.   The last example is seen at $z=0.1$ in the bottom 
right
corner.  The circle encloses S13, a satellite that is also disturbed as it enters this hot medium.  The long tail seen diagonally extending to the right is likely a product of ram pressure stripping.  

We follow the evolution of the settled gas within the dark matter halo of the satellites as they enter our projection map volume. When necessary, we examine the stellar mass distribution maps to locate the exact positions of the satellites. We are able to identify and track the gas inside the satellite cores since it is compact
and does not mix easily with the 
rest of the
gas in the halo.  We start tracking the gas content from the moment 
that the satellites
appear in the box analyzed until they lose all of their gas, exit the simulation box, merge with the main disk or until $z=0$.  The total amount of gas inside the satellites is shown in Figure~\ref{fig:evolution} by the dotted line, and is calculated at every five simulation outputs.  As seen in the curve, there is more satellite gas at earlier times, showing that there is a greater number of satellites at around $z=0.5$, with many of 
these, such as S3, being gas rich.
\begin{deluxetable*}{cccccccccc}
\tablenum{1}
\tablecolumns{9}
\tablewidth{0pc}
\tablecaption{Mass Loss Rates}
\tablehead{
\colhead{Satellite}&
\colhead{$z_0$}& 
\colhead{$z_f$} & 
\colhead{Time} & 
\colhead{$M_o$}&
\colhead{$M_f$}&
\colhead{$\Delta M$}&
\colhead{$\Delta \dot{M}$}&
\colhead{$D$}&
\colhead{Mass }\\
\colhead{Number}&
\colhead{}&
\colhead{}&
\colhead{(Myr)}&
\colhead{($10^6 M_{\odot}$)}&
\colhead{($10^6 M_{\odot}$)}&
\colhead{($10^6 M_{\odot}$)}&
\colhead{($M_{\odot}$ yr$^{-1}$)}&
\colhead{(kpc)}&
\colhead{Loss}}

\startdata 
 S1  & 0.495 & 0.221 &  2346 & 0.07     & 0.01     & 0.06      &  $2.56 \times 10^{-5}$ & 45  & 86\%\\
 S2  & 0.495 & 0.300 & 1579 & 3.66     & 0.31      & 3.35      &  $2.12 \times 10^{-3}$ & 66  & 92\%\\
 S3  & 0.489 & 0.339 & 1195 & 174.00 & 125.92 &48.08  &  $4.02  \times 10^{-2} $ &  0    & 28\% \\
 S4  & 0.477 & 0.443 &  256  & 87.25 & 46.23   &  41.02   &  $1.60 \times 10^{-1}$   & 31   & 47\% \\
 S5  & 0.454 & 0.339 &  939  & 30.75   & 1.31      &  29.44   &  $3.14 \times 10^{-2}$ & 57  & 96\%\\
 S6  & 0.374 & 0.282 &  811  & 1.92      & 0.64      &  1.28    &   $1.58 \times 10^{-3}$ &  0  & 67\%\\
 S7  & 0.363 & 0.282 &  725  & 22.09   & 12.46    &  9.63  &   $1.33 \times 10^{-2}$    &  0 & 44\% \\
 S8  & 0.363 & 0.255 &  981  & 20.78  &  27.28  & $-6.50$ &  $-6.63 \times 10^{-3}$  & 0 &$-$31\%\\
 S9  & 0.287 & 0.238 &  469 & 2.20      & 0.01      &  2.19      &  $4.67 \times  10^{-3}$ & 0 & 99\%\\
 S10 & 0.334 & 0        & 3712 & 46.25   & 37.25   & 9.00    & $2.42 \times 10^{-3}$     & 132  & 19\%\\
 S11 & 0.247 & 0        & 2901 & 40.80   & 12.58   & 28.22  & $9.73 \times 10^{-3}  $   & 45 & 69\%\\
 S12 & 0.221 & 0.033 & 2176 & 8.14 &  0.13     & 8.01   & $3.68 \times 10^{-3}$       & 68   & 98\% \\
 S13 & 0.196 & 0        & 2389  &  53.14 & 0.71     & 52.43    & $2.19 \times 10^{-2}$  &  37   & 99\%\\
 S14 & 0.196 & 0.172 & 256   & 0.10    & 0.01      & 0.09     & $3.52  \times 10^{-4}$  & 67  & 90\%\\
 S15 & 0.196 & 0         & 2389 & 19.22  & 15.38   & 3.84   & $1.61  \times 10^{-3}$    & 28 & 20\%\\
 S16 & 0.196 & 0.184 & 128   & 0.03 & 0.01         & 0.02     & $1.56  \times 10^{-4}$   &  81  & 67\%\\
 S17 & 0.196 & 0.149 & 512    & 0.05 & 0.05        & 0       & 0                                     &  282 & 0\\
 S18 & 0.180 & 0.024 & 1877 & 0.60 & 0.01         &  0.59    & $3.14  \times 10^{-4}$    & 64  & 98\%\\
 S19 & 0.105 & 0         & 1365 & 17.54 & 2.42       & 15.12   & $1.11  \times 10^{-2}$    & 0 & 86\%\\
 S20 & 0.084 & 0.056 & 341  & 0.03 & 0.01         & 0.02      & $5.87  \times 10^{-5}$     & 248 & 67\%\\
\enddata

\end{deluxetable*}

Table 1 summarizes our satellite analysis. 
It 
lists the satellite IDs, initial and final redshifts, number of years this redshift interval corresponds to, initial and final masses, mass losses, mean mass loss rates, projected distances of closest approach, and the percentage of mass loss.  We see a variety of satellites with a range of initial HI content  $10^4 -10^8$ M$_\odot$ that generally make a close encounter with the main galaxy before losing all of their gas or exiting the simulation box. The majority of these galaxies are losing their HI content at different rates.  Thirteen satellites lose 50\% or more of their original gas content, with nine of these losing 80\% or more.   The other satellites have lost less than 50\% of their initial HI content (5), one kept its gas content, and one actually gained some HI due to 
an
interaction with companion satellites.  

The satellite HI mass loss rate is not necessarily correlated with the projected distance of closest approach.  The two satellites with largest distances ($d > 200$ kpc) have mass loss rates that are negligible.  At $d<200$ kpc, the mass loss rate for each satellite varies since it depends on the initial HI mass of the satellite, the 3D location, and the timescale at which 
it 
is losing gas.  For example, S1 is not very massive in HI and loses gas slowly, which leads to an insignificant rate.  On the other hand,  S4 is fairly massive and loses its gas on a very short timescale, translating to the highest mass loss rate.  
The median mean mass loss rate for all of the satellites that have lost HI gas is $3.1 \times 10^{-3}$ M$_\odot$ yr$^{-1}$.   If we calculate the median mean mass loss rate for only the galaxies that have gas at $z=0$, we get $9.7 \times 10^{-3}$ M$_\odot$ yr$^{-1}$.    In a Gyr, these numbers would translate to  $3.1 \times 10^{6}$ M$_\odot$
and  $9.7 \times 10^{6}$ M$_\odot$ of stripped gas, showing that satellite gas provides an important, but not dominant, origin for the cold halo gas. 

There is the question, however, of how long the stripped neutral gas will remain in the halo before it gets disrupted or ionized.  It is difficult to make a general estimate on this timescale since it depends on unique conditions such as 
the 
HI mass, the properties of the medium, and the velocity.   We find that the stripped gas in the simulation survives for 2--5 outputs, which corresponds to 80--200 Myr.  This agrees with previous detailed studies \citep[e.g.,][] {Mayer06} showing that stripped gas can survive for 100 Myr.  

There are various possible gas-loss mechanisms including tidal interactions, ram pressure stripping, star formation, or supernovae feedback.  Two satellites, S13 and S19, are prime examples for ram pressure stripping. We have checked the stellar components of the satellites, and they show no evidence of interaction, implying that tidal interactions are minor. In addition to this,  we have constructed star formation histories for the five satellites that still have gas at $z=0$; none shows a recent burst of star formation, indicating the HI gas was stripped rather than consumed.

\subsubsection{Streams}
\begin{figure*}
\begin{center}
\includegraphics[trim= 10 70 0 70,clip,scale=0.8]{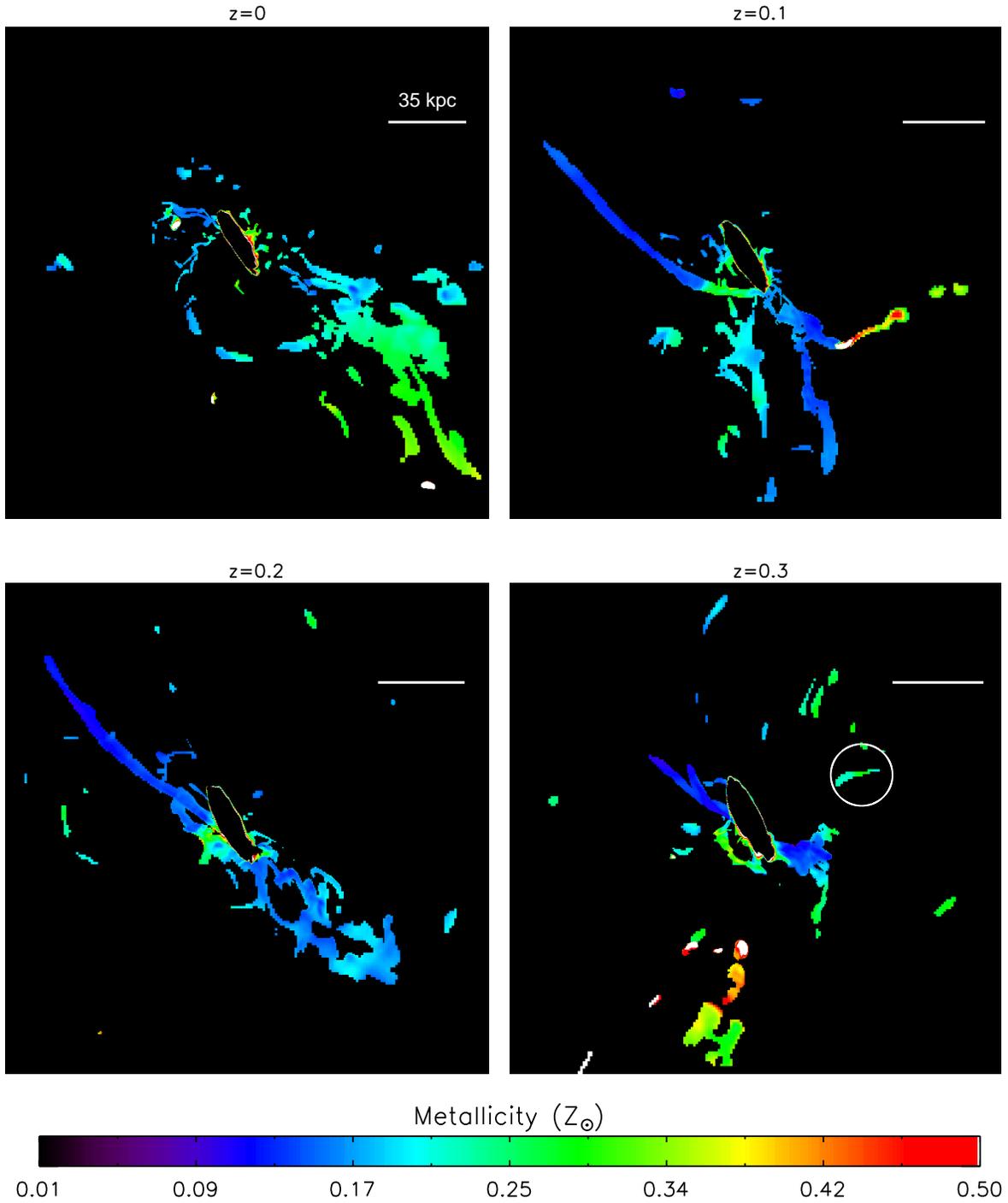}
\caption{Metallicity maps at different redshifts.  The plotted range is from 0.01 to 0.50 Z$_\odot$ for contrast, but satellite gas generally has higher metallicities which appear as white regions.  The bar in the upper right corner in each map corresponds to 35 kpc.  Here we only include cells that have at least $10^{16}$ cm$^{-2}$ in HI column density.}
\label{fig:metMap}
\end{center}
\end{figure*}

In contrast to our satellite analysis, we cannot track the filaments accurately since they are not tightly bound and consequently do not have a defined structure.  They fragment easily and interact with other gas components present in the halo.  This limits a precise quantitative assessment of the amount of cold flow gas present at different times.  We can, however, qualitatively describe what 
happens
at different redshifts using the 
movie, snapshots of which 
are shown in Figure~\ref{fig:HIMaps}.  At $z=0.5$, the gas in the vicinity of the disk has not settled, and there are three main filaments that are directly feeding the disk, accompanied by various smaller and less dense structures.  The main filaments are very diffuse and tend to remain in the same directions (upper-left, upper-right, and lower-center). The small-scale filaments are dominant in the second snapshot, where they are still delivering gas to the disk.   Some of these streams are intermittent and appear to be inactive at certain times, but then become a prominent feature again. As time progresses, the filaments tend to fragment, become less dense and are unable to directly reach the disk as coherent structures.   For example, at $z=0.1$, there is a dominant filamentary stream to the left of the simulated galaxy with an HI mass of $\sim 3 \times 10^7 ~\rm M_\odot$ .  This stream moves around the galaxy in the counterclockwise direction, and collides with a smaller stream that is entering from the bottom of the simulation box, fragmenting the flow and leaving several discrete features concentrated in the bottom-right corner of the simulation box at $z=0$ with an approximate total mass of  $4 \times 10^7 ~\rm M_\odot$. It is important to note that we do not expect numerical fragmentation since all cells inside filaments satisfy the Truelove criterion.
We look at the metallicity of the gas as a possible way to place a constraint on the amount of cold flow gas at different times.   In principle, the satellite gas should have a higher metallicity than the gas due to filamentary streams.  Figure~\ref{fig:metMap} shows metallicity maps at different redshifts.  These are projected metallicities and are density-weighted along each line-of-sight. The disk has been removed and we only plot metallicities for cells that have at least $10^{16}$ cm$^{-2}$ in HI column density.   We restrict our metallicity range to $0.01-0.50$ Z$_\odot$ for better contrast.  Many satellites have metallicities above 0.5 Z$_\odot$, which appear white in the metallicity map.  As expected, we do see a difference in metallicity between satellite and filamentary stream gas.  The best example of this is seen at $z=0.1$, where S13 is seen getting stripped in the middle-right of the plot, and where various cold streams are approaching the disk.  The metallicity of the streams is around 0.2 Z$_\odot$, and the stripped gas from the satellites generally has values above 0.3 Z$_\odot$.  

Despite the distinction here, there is 
also mixing going on, which is mostly due to supernova feedback from the galactic disk interacting with the filaments.  At $z=0$, the gas in the bottom-right corner of the plot originates from filamentary streams contaminated by the supernova wind, as shown by the temperature and metallicity movies.   Another example of contamination is seen in the fourth panel ($z=0.3$), where the elongated features with higher metallicities above the disk are not associated with satellite gas.  Some of these features condensed out of filamentary streams but were contaminated by the higher-metallicity supernova wind.  The feature enclosed by the white circle is a product of mixing between cold flow material and a fragmented supernova wind shell.  

We conclude from this analysis that all lower metallicity gas ($\lesssim$ 0.2 Z$_\odot$) corresponds to cold flows in the simulation, but not all higher metallicity gas is due to satellites.  This allows us to place a minimum on the amount of gas that is due to cold flows at different redshifts.  The dashed line in Figure~\ref{fig:evolution} represents the total amount of HI at each redshift with metallicities less than  0.2 Z$_\odot$.  This is a minimum since the amount of gas originating from cold streams is higher, but some of it is indistinguishable from satellite gas in terms of metallicity. The peaks in this curve roughly correspond to the maxima in the solid curve, showing that active cold flows dominate the HI content at those particular times.  The relative amount of this low-metallicity gas with respect to the total amount of HI halo gas at different redshifts ranges between 25\% and 75\%.

\section{Comparison with Observations}

\subsection{Andromeda}
\begin{figure}
\begin{center}
\includegraphics[trim= 0 0 100 5,clip,scale=0.65]{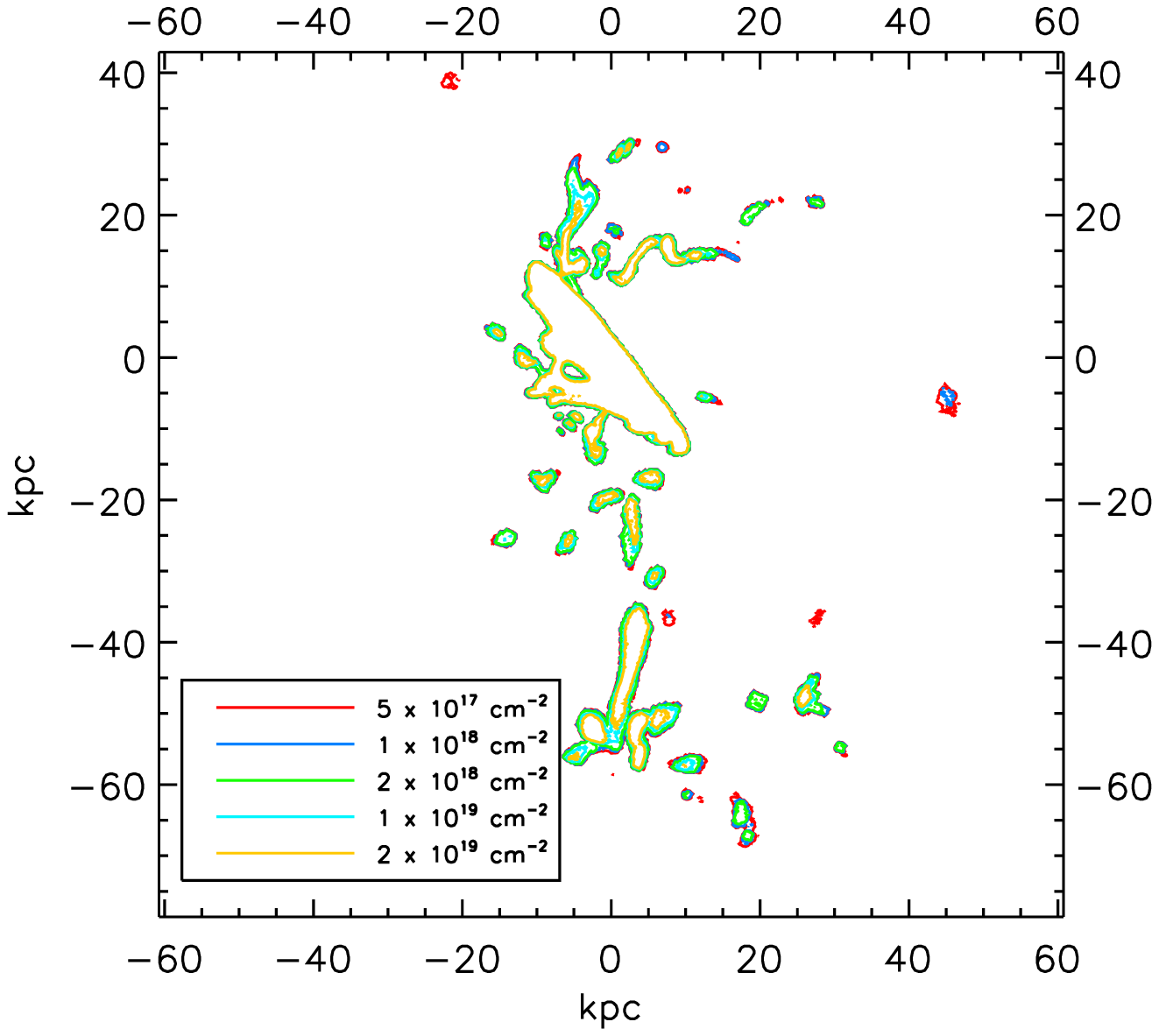}
\caption{Total HI distribution of the simulated galaxy at $z=0$ where we have rotated the galaxy to M31's inclination ($i = 78^\circ$) and position angle (PA=$35^\circ$),  Contours  are drawn in levels of (0.5, 1, 2, 10, 20) $\times 10^{18}$ cm$^{-2}$, the same levels as T04.  We color-code the contours according to the legend in the plot.}
\label{fig:M31}
\end{center}
\end{figure}

We first compare the simulation with the deep HI observations of M31 carried out by \citet[][hereafter T04]{Thilker04}.  We choose M31 since it is the nearest spiral galaxy of similar size and mass to the Milky-Way and several properties of the clouds can be directly compared to the simulation.   The wide-field HI observations (94 $\times 94 $ kpc$^{2}$) were conducted with the Green Bank Telescope (GBT) and achieve a rms flux sensitivity of 7.2 mJy beam$^{-1}$ (1$\sigma$), which  corresponds to a  column density of $2.5 \times 10^{17}$ cm$^{-2}$ over 18 km s$^{-1}$.   The data include velocities between $-$827 and 226 km s$^{-1}$, where $-$300 km s$^{-1}$ is the systemic velocity of the system.  The authors exclude data with velocities more positive than $-$160 km s$^{-1}$ due to Galactic emission.  After removing the disk HI emission, they find at least 20 sources in the halo above $2\sigma$ within 50 kpc of the disk and with a total HI mass of $3 \times 10^7$ M$_\odot$.  Follow-up observations carried out at Effelsberg by \cite{Westmeier08} indicate there are no clouds above a $3\sigma$ mass detection limit of  $8 \times 10^{4}$ M$_\odot$ at distances greater than 50 kpc from the disk.

Figure~\ref{fig:M31} shows the total HI distribution of our simulated galaxy at $z=0$ matching the viewing geometry of M31. We have rotated the galaxy accordingly to correspond to M31's inclination ($i = 78^\circ$) and position angle (PA=$35^\circ$), and use the same contour levels as T04 [(0.5, 1, 2, 10, 20) $\times 10^{18}$ cm$^{-2}$]. If we exclude the gas in the disk, we find a total HI mass of $1.1 \times 10^8$ M$_\odot$. We detect about 30 discrete sources, and many are arranged in complexes. Some of the structures are compact clouds with masses within the range of $10^4 - 10^6$ M$_\odot$, while others have an elongated filamentary morphology.  These filamentary features have higher masses, such as the nearly horizontal structure located about 15 kpc above the disk, that has a HI mass of $1.1 \times 10^7$ M$_\odot$.

The morphology, distribution and kinematics of the clouds are similar to the observations. T04 detect a wide array of morphological features with similar masses to the ones seen in the simulation.  The cold halo gas in the simulation does seem to extend beyond 50 kpc from the disk, as is seen in the large complex south of the disk that extends to about 70 kpc.  If we limit the HI to only 50 kpc to better compare with observations, we detect about 88\% of the value quoted above, which means that the bulk of the gas is located within 50 kpc of the galaxy.   One issue of concern we see in this comparison is that our contour levels are closely spaced together, showing that our simulation is over-predicting values for the column density for some of the clouds. This discrepancy might be a consequence of not including interstellar UV in the simulation \citep{BH02,MaloneyPutman03}.   Kinematically, the comparison between the GBT data and the simulation is more complicated since the observations mostly probe redshifted velocities to avoid Galactic emission.  In general, T04 find that the discrete features have velocities in between $-515$ and $-172$  km s$^{-1}$, which can be compared to M31's $V_{sys}=-300$  km s$^{-1}$. In contrast, M31's HI gas with stream-like morphology has values within $\sim80$ km s$^{-1}$ of the systemic velocity.  This is roughly consistent with the simulated HI halo gas, where many of the diffuse streams are close to the systemic velocity and the compact features have a wider range in velocities.   We also compare the covering fraction of M31 with the simulation in \S4.3.

\subsection{Other Galaxies}
The population of high-velocity clouds in the Milky Way has been extensively studied over the past decades.  We can compare the distribution, amount, and metallicities of HVCs to the cold halo gas seen in the simulation. As various surveys have shown \citep[e.g.,][]{Putman02,Kalberla05,Peek11}, the HVCs are grouped into large complexes, which is consistent with the distribution of the gas in the simulation.  If we exclude the gas in the Magellanic Stream, and place all of the HVCs at a distance of 10 kpc, there is a total of $\sim 3 \times 10^7$ M$_\odot$ in HI (Putman et al. 2012).  In addition to this, HVCs in the Milky Way exhibit metallicities at around $\sim 0.1-0.3$ Z$_\odot$ \citep{Wakker01}, which is consistent with the values of the simulated HI gas.  

In addition to the Milky Way and M31, other spirals have been shown to have cold gas in their halos.  Another well studied galaxy is NGC 891  \citep[e.g.,][]{Oosterloo07}, which has been shown to have an extended and massive halo of HI with a mass of $1.2 \times 10^9$ M$_\odot$ when both the halo and disk-halo interface region are examined, accounting for about 30\% of the total mass of neutral gas found in the galaxy.  Its distribution of cold gas is not smooth, showing clouds with HI masses of $\sim 1 \times 10^6$ M$_\odot$, and an extended filament extending to the northwest of the galaxy. If the disk and disk-halo interface region are subtracted from the HI distribution map, the remaining neutral gas is  $\sim 1 \times 10^8$ M$_\odot$ in the form of clouds and the extended filament \citep{Sancisi08}, resembling what is seen in the simulation.  

\citet{Sancisi08} quotes that $\sim 25$\% of field galaxies have extra-planar features of $\gtrsim$ $10^8$ M$_\odot$. This number would go up to 50\% if warps and kinematic deviations from the disk are included \citep{Sancisi08,Haynes98}. Ongoing deep HI observations \citep[e.g., HALOGAS][]{Heald11} are further examining the cold halo gas of a larger sample of spirals.  We expect these will only detect part of the features seen in the simulation. First, the column densities being probed in the observations are higher ($>10^{19}$ cm$^{-2}$) than the bulk of the simulated clouds.  In addition to this, the average column densities in the simulation may be lower since we did not include interstellar UV.  Future HI deep surveys will allow us to compare the simulation with a larger sample of galaxies. 

\subsection{Covering Fraction}
\begin{figure}
\begin{center}
\includegraphics[trim= 45 10 10 5,clip,scale=0.55]{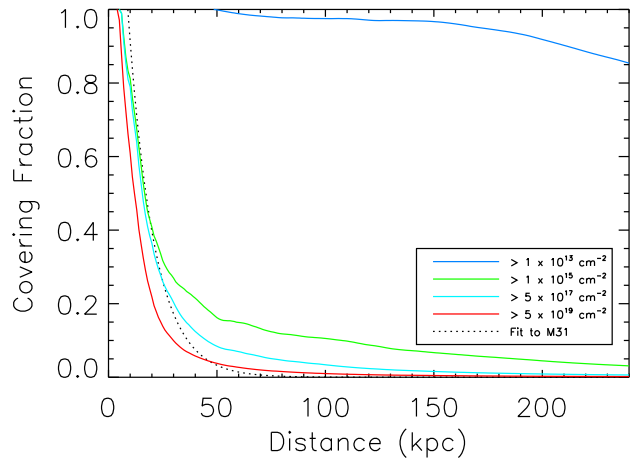}
\caption{Covering fraction as a function of distance for various column density limits calculated using the level 8 extraction box.  An exponential fit to the M31 halo population is included for comparison (shown in black; \citet{Richter11}).  We color-code the curves according to the column density, as shown in the legend in the plot.}
\label{fig:covering}
\end{center}
\end{figure}

We calculate the projected covering fraction at $z=0$ as a function of distance for various column density limits (shown in Figure~\ref{fig:covering}).  We use the level 8 extraction box to calculate the covering fraction out to 240 kpc for different column densities. Here we plot four representative values: $1\times 10^{13},~ 1 \times 10^{15},~ 5 \times 10^{17},~ 5 \times 10^{19}$ cm$^{-2}$.  We also include a curve showing the exponential fit  of $f=2.1\exp(-d/12)$ done by \citet{Richter11} to the covering fraction for M31. 

Most of the curves show the covering fraction exponentially decreasing as a function of distance.  We calculated covering fractions for intermediate values between $10^{13}$ and $10^{15}$  cm$^{-2}$, and found the exponential drop-off to hold true for N$_{HI} > 5 \times 10^{13}$ cm$^{-2}$.  Below this value (shown by the blue curve), we find covering fractions near unity out to 50 kpc, with the value slowly decreasing afterwards to a minimum of 0.85 at 240 kpc.  This indicates that the halo of the simulated galaxy is filled with low-density gas.  In contrast, the covering fraction for N$_{HI} > 1 \times 10^{15}$ cm$^{-2}$ is less than 0.2 beyond 50 kpc, indicating that denser absorbers are not a common feature in the extended halo.  

The calculated covering fractions agree well with absorption-line work and HI observations.  There are several studies \citep{Bowen02,Wakker09, Prochaska11} that have detected Ly$\alpha$ absorption against background quasars, probing low-density gas in the range $10^{15}>$ N$_{HI} > 10^{13}$ cm$^{-2}$.  All of these studies find covering fractions close to 100\% within 300 kpc of a galaxy, which agrees with the blue curve.  For higher-density gas,  we can compare our  results to the covering fractions from the HI data of M31.  The exponential fit found by \citet{Richter11} expresses the covering fraction as a function of distance.  If we compare this fit to the cyan curve (N$_{HI} > 5 \times 10^{17}$ cm$^{-2}$), which is the lowest contour value in T04, we find similar values for the covering fraction.   
The observational fit to larger distances is not reliable since there are no detections at $d>60$ kpc.  The covering fraction of HVCs in the Milky Way above $7 \times 10^{17}$ cm$^{-2}$ is 0.37 \citep{Murphy95}, which is roughly consistent with the cyan curve at $\sim30$ kpc.

\section{On Feeding a Spiral Galaxy}
Our study addresses how spiral galaxies get their fuel at the present time.  The constant star formation rate and stellar metallicity measurements indicate that galaxies must be constantly accreting a fresh supply of gas (see \S 1).  The simulation shows that the galaxy has had a nearly constant supply of atomic gas in the halo ($\sim 10^{8}$ M$_\odot$) since $z=0.3$.   At $z>0.3$ the supply of gas was higher, which is consistent with previous work by \citet{KeresKatz05}, where they show that cold flows are able to directly feed the disk at higher redshift.  At low redshift, we see that the filaments are no longer able to directly reach the disk, and fragment due to instabilities in the halo.  

The next important issue is to determine if the accreting neutral gas will be enough to maintain the observed star formation rate.  Our calculated value of $\dot{M} \sim 0.2$ M$_{\odot}$  yr$^{-1}$ (see \S 3.2) is consistent with the HI observations of the Milky Way \citep{Putman12}.  This indicates that there must be other sources for continued star formation given the galaxy's SFR of $1.9$  M$_{\odot}$  yr$^{-1}$ \citep{Chomiuk11}.  Recent observations suggest that some of the missing gas might be ionized \citep{Shull09,Lehner11}, 
which is consistent with detailed simulations of halo clouds \citep{Bland07,HeitschPutman09}.  
A preliminary calculation for the total gas mass accretion rate from the simulation in this study yields a value of $\sim 3$ M$_{\odot}$ yr$^{-1}$, indicating the bulk of the accretion is not detectable in HI (Joung et al. 2012).

The origin of cold halo gas has been extensively studied over the past decade \citep[e.g.,][]{Maller04,Keres09}. Our study shows that the clouds come from stripped satellites and filamentary material.  These two cloud formation mechanisms are evident at all times, where we see satellites getting stripped by the hot gas, and filaments entering the halo.  In some instances the filaments are able to get close to the galaxy (within $\sim20$ kpc) before getting disrupted or colliding with other streams. In other cases, the filaments do not get close to the halo, but some of the material is able to cool down at larger distances and form compact clouds.  This is consistent with \citet{Joung11}, where they show that a nonlinear perturbation such as filamentary streams is needed to seed the formation of clouds in the hot medium.  Even though satellites are losing their majority of the HI, the stripped cold gas has a relatively short timescale.   We find that stripped gas from satellites is an important but not a major constituent of the cold halo gas.   Filamentary material will also have a similar timescale but is more abundant and is being more consistently replenished, thus usually having a more dominant role. As seen in Figure~\ref{fig:evolution}, the low-metallicity gas can account on average for 54\% (ranges from 25\% to 75\%) of the HI halo gas seen in the studied redshift interval.   In addition to these two cloud formation mechanisms, we find that in some instances the supernova shell fragments, mixes in with filamentary flows and then falls back to the disk. 

Distinguishing between the two main cloud formation mechanisms is challenging since we do not have the capability of tracking the origin of each gas particle, such as in SPH simulations.  Using metallicity as a tracer allows us to place constraints on the amount of cold halo gas that is due to filamentary streams.  As seen in Figure~\ref{fig:metMap}, the metallicity of satellite gas has values above  0.3 Z$_\odot$, while the cold flows have metallicities of $\sim0.1$ Z$_\odot$.  We find that the metallicity of the cold halo gas at $z=0$ has values between 0.01-0.5 Z$_\odot$, with most of the gas showing some degree of enrichment, including the cold flows.  At higher redshift, the filaments in the simulation have lower metallicities .  Recent work by \citet{Fumagalli11} finds a mean metallicity of 0.01 Z$_\odot$ for cold streams at $z>1$, which is consistent with our results, and the restrictions from the G-dwarfs (see references in \S1).   The mixing at lower redshift indicates one should be cautious when trying to use metallicities as a diagnostic for the origin since the high-metallicity gas could at least partly originate from cold streams that have been enriched by galactic feedback processes. 

The properties of halo gas around galaxies have only begun to be explored in detail with observations and simulations.  In HI emission, there is only a handful of galaxies whose haloes have been mapped.   The next generation of radio telescopes  (e.g., EVLA, ASKAP, MeerKat, SKA) will enable us to reach to lower column densities to characterize the frequency and properties of the neutral halo gas features in a larger sample of galaxies. In absorption, the Cosmic Origins Spectrograph on HST is probing the diffuse neutral gas and the warm-hot ionized gas in a large sample of galaxies \citep{Tumlinson11}.  On the simulation front, significant progress has been made in recent years in achieving high resolution and including much of the relevant physics.  It is now necessary to carry out high resolution simulations of galaxies of different sizes and in different environments to fully understand the origin and fate of halo gas in a wider range of galaxies.  Both approaches will enable us to understand the origin, state and future of the gas in the halo of galaxies.

\section{Conclusion}
In this paper, we studied how spiral galaxies such as our Milky Way get their gas.  To do this, we use a high resolution cosmological grid simulation to study the cold halo gas in the redshift interval $0.5 > z >0$.  We summarize our results below:
\begin{itemize}
 
\item The cold halo gas seen in the simulation at $z=0$ is consistent with observations of M31 and the Milky Way.  We find $\sim 10^8$ M$_\odot$ of atomic gas  in the halo, and features that are mostly arranged into complexes of clouds with spherical or stream-like morphology. The majority of the simulation clouds have velocities that are consistent with infall or have values near the systemic velocity.

\item There is a nearly constant amount of HI halo gas since $z=0.3$ with an increasing amount of HI with decreasing galactocentric radius. This suggests that Milky Way-sized  galaxies have had a constant supply of star formation fuel. 

\item We calculate a HI mass accretion rate of  $\dot{M} \sim 0.2$ M$_{\odot}$  yr$^{-1}$ indicating the need for additional sources to sustain the current star formation rate.  \\

\item We calculate the covering fraction of cold halo gas as a function of radius at $z=0$ for different column densities.  We find an exponential drop-off for N$_{HI} > 1 \times 10^{15}$ cm$^{-2}$ within the virial radius and a nearly unity covering fraction throughout for N$_{HI} > 1 \times 10^{13}$ cm$^{-2}$.  These values are consistent with what is found for M31, the Milky Way, and those reported for Ly$\alpha$ absorbers.

\item The HI halo gas originates primarily from filamentary streams and stripped satellites.  To a lesser extent, we see evidence of supernova shells mixing with filamentary material and fragmenting and cooling. 

\item We identify 20 satellites in the redshift interval $0.5 > z > 0$ that have gas.  Fifteen of these lose 50\% or more of its original gas content as they enter the hot medium.  The median mean mass loss rate is   $3.1 \times 10^{-3}$ M$_\odot$ yr$^{-1}$.

\item Filaments are seen at all redshifts, being able to reach the disk at $z \gtrsim 0.5$ but fragmenting at lower redshifts.  Some are intermittent and become more active at particular times, resulting in peaks in the HI halo gas mass.  

\item We use the metallicity of the gas to distinguish between satellite gas and cold flow material. We find that filamentary streams account for at least 25-75\% of the cold halo gas at a given redshift. The metallicity analysis also indicates that supernova winds can contaminate the HI gas originating from filamentary streams.  

\end{itemize}

We acknowledge support from NSF grants AST-1008134 and AST-0904059, and the Luce Foundation.  We thank Josh Peek for help throughout the project, and Jacqueline van Gorkom, Kathryn Johnston and Greg Bryan for useful discussions.

\bibliographystyle{apj}
\setlength{\baselineskip}{0.6\baselineskip}
\bibliography{references}
\setlength{\baselineskip}{1.667\baselineskip}

\end{document}